\documentclass[aps,prd,reprint,preprintnumbers,superscriptaddress,nofootinbib,floatfix]{revtex4-1}

\usepackage{amsmath,amssymb,amsfonts,dcolumn,color,graphicx,graphics,latexsym,placeins,epsfig}
\usepackage{epsfig}
\usepackage{bm}
\usepackage{slashed}
\usepackage{latexsym}
\usepackage{natbib}
\usepackage{url}
\usepackage{dcolumn}
\usepackage{color}
\usepackage{amsfonts,amssymb,amsmath}
\usepackage{graphicx,epsfig}
\usepackage{psfrag}
\usepackage{subfigure}
\usepackage{tabularx}
\usepackage{hyperref}
\hypersetup{colorlinks=true}
\usepackage{comment}

\newcommand{\be}{\begin{equation}}
\newcommand{\ee}{\end{equation}}
\newcommand{\ba}{\begin{eqnarray}}
\newcommand{\ea}{\end{eqnarray}}

\begin{document}
  
\preprint{IPPP/26/35}
  
\title{Searching for ultralight bosons with Josephson junction interferometry}

\newcommand{\IPPP}{\affiliation{Institute for Particle Physics Phenomenology (IPPP), Department of Physics, Durham University, Durham DH1 3LE, United Kingdom}}

\author{Djuna Croon}
\email{djuna.l.croon@durham.ac.uk}
\IPPP
\author{Tanmay Kumar Poddar}
\email{tanmay.k.poddar@durham.ac.uk}
\IPPP

\begin{abstract}
Ultralight bosons sourced by macroscopic objects can generate long-range spin-independent and spin-dependent potentials that are accessible to precision interferometry. Such potentials induce phase shifts in Josephson junctions, detectable through precision current measurements. We propose three experimental scenarios to probe photophilic scalar interactions, Lorentz-violating scalar-mediated interactions, and axion-mediated monopole-dipole interactions, depending on the nature (unpolarized or polarized) of the source. The proposed setups provide sensitivities to novel mixed couplings that are largely unconstrained by existing bounds and enables the exploration of new forces at centimeter to micrometer length scales.
\end{abstract}

\maketitle
\section{Introduction}

Ultralight bosons--such as scalar fields \cite{Morisaki:2018htj,Brzeminski:2022sde}, pseudoscalar axions \cite{Peccei:1977hh,Weinberg:1977ma}, and more generally particles of higher spin \cite{KumarPoddar:2019ceq,Cembranos:2017vgi} can be sourced by macroscopic objects provided they feebly couple to Standard Model (SM) fields, as permitted by current experimental constraints \cite{Moody:1984ba}. The properties of such interactions are determined by two parameters--the coupling strength to SM particles and the interaction range. The latter is set by the boson Compton wavelength, $\lambda_c= m_\phi^{-1}$, where $m_\phi$ denotes the mass of the boson $\phi$, and therefore becomes macroscopic for sufficiently small masses. The interaction strengths of such ultralight particles are constrained from several experiments \cite{AxionLimits}. The required couplings must be extremely weak, otherwise deviations from known forces would already have been observed experimentally \cite{Fischbach1992,Adelberger:2009zz}. Owing to their light masses and stability, these particles also constitute well motivated dark matter (DM) candidates and can account for a significant fraction of the observed cosmological DM abundance \cite{Hu:2000ke,Hui:2016ltb}.

The ultralight bosonic interactions sourced by macroscopic bodies composed of non-relativistic fermions generically gives rise to a potential of Yukawa-type or its derivative \cite{Moody:1984ba}. The phenomenology of the resulting interaction depends on the operator structure of the boson-matter coupling: dilaton-like couplings produce spin-independent forces, whereas pseudoscalar couplings to fermion axial currents produce spin-dependent long-range interactions. These interactions may give rise to macroscopic or fifth forces \cite{Moody:1984ba,OHare:2020wah,Cong:2024qly}, which can be tested through high-precision tests of gravity. 
Such tests are available over a wide range of scales, from galactic and cosmological distances \cite{Archidiacono:2022iuu,Tsai:2021irw,Tsai:2023zza,Poddar:2021sbc,Hook:2017psm,Lambiase:2025qyl,Carenza:2025jwn,Poddar:2023pfj,KumarPoddar:2020kdz,KumarPoddar:2019ceq,KumarPoddar:2019jxe} down to laboratory separations of order centimeters to millimeters \cite{Adelberger:2009zz,Wagner:2012ui}. At shorter distances, however, experimental tests remain comparatively limited, leaving open the possibility of deviations from the Newtonian potential. Such modifications could arise from the exchange of ultralight bosons between nearby objects. As the mediator mass ranges $m_\phi\sim \mathcal{O}(10^{-6}~\mathrm{eV}-10^{-2}~\mathrm{eV})$, the corresponding force range spans as $\lambda_c\sim \mathcal{O}(10~\mathrm{cm}-10~\mu \mathrm{m})$. Beyond this length scale, the force becomes short-ranged and the interaction becomes exponentially suppressed. Consequently, torsion-balance and related precision gravitational experiments progressively lose sensitivity in this mass regime.

A number of experimental strategies have been proposed to probe light bosonic fields at short distances using condensed-matter systems and emerging quantum technologies \cite{Piegsa:2012th,DeMille:2017azs,Chu:2021agg,Wang:2022bpw,Cong:2024qly,Gao:2025ryi,Sheng:2025ygn,Poddar:2026hch}. Among these, Josephson-junction (JJ)-based detectors have attracted particular interest. Such setups have been shown to be sensitive, for example, to $U(1)_{B-L}$ vector gauge bosons, whose coupling to fermion number produces a spin-independent macroscopic force \cite{Cheng:2024yrn}. Related configurations have also been explored in follow-up studies aimed at testing the quantum aspects of gravity at millimeter scales \cite{Cheng:2024zde}.

In this paper we build on these results by considering JJ-based setups with different source configurations, generating both spin-independent and spin-dependent forces mediated by scalars and pseudoscalars in the mass range $m_\phi\sim \mathcal{O}(10^{-6}-1)~\mathrm{eV}$. 
Specifically, we consider electron-scalar-photon mixed couplings, Lorentz-violating scalar-mediated interactions in the electronic sector, and axion-mediated monopole-dipole couplings involving electrons. By incorporating an appropriate readout scheme together with noise characterization and background reduction strategies, we derive new projected sensitivities to these mixed couplings using a single JJ-based experiment operating at centimeter to micrometer-scale separations, a regime that has not previously been explored in a unified framework.
The proposed setups are designed to operate with minimal backgrounds and controlled noise, enabling a direct and robust search in this parameter space.

\section{Josephson Junction Response to ultralight boson-Induced Potentials}

We briefly outline here the general response of a JJ to an ultralight boson-induced potential, before focusing on specific examples. In this work, we consider superconductor-insulator-superconductor (S–I–S) JJ consisting of two superconducting electrodes separated by an insulating barrier \cite{JOSEPHSON1962251,RevModPhys.46.251}. 
Each electrode hosts a macroscopic coherent state of Cooper pairs, characterized by a well-defined phase \cite{Ginzburg2009}. We denote the characteristic size of each electrode by $a$, and the separation between the two superconducting regions by $\delta \sim a$. An external source of non-relativistic SM particles located at a distance $d$ generates a scalar field profile, which induces a position-dependent potential $V_\phi(r)$ acting on the Cooper-pair electrons.

The phase accumulated by a Cooper pair in a superconducting electrode located at a position $r$ over a coherent integration time $T$ is given by\footnote{We use natural units with $\hbar=c=1$, unless stated otherwise.}
\begin{equation}
\varphi(r,T) = -\int_0^T V_\phi(r)\, d\tau,
\end{equation}
which follows from the time-dependent Schr\"odinger equation $i\partial_t \psi = H \psi$ for $\psi=\sqrt{n} e^{i\varphi}$, where $n$ denotes the number density of the Cooper pairs in the superconducting electrode. Although the scalar-induced potential varies across each superconducting electrode, the Cooper pairs within a given electrode share a single macroscopic phase. Any spatial phase gradient would generate a supercurrent and is therefore energetically suppressed. Instead, small adjustments in the local Cooper pair density maintain phase coherence throughout the electrode. As a result, each superconducting lead can be described by a single collective phase, which effectively corresponds to an average over microscopic positions. The phase difference $\Delta \varphi$ is therefore controlled by the difference in the scalar potential evaluated at the two electrodes.

Since the scalar-induced potential depends on the distance to the source, the two superconducting electrodes, located at positions separated by $\delta$, generally acquire different phases. The physically relevant quantity is the phase difference
\begin{equation}
\Delta \varphi = \varphi_1(r_1, T) - \varphi_2(r_2, T),
\label{difft}
\end{equation}
which depends on the spatial variation of $V_\phi(r)$ across the junction. This phase difference gives rise to a measurable current. The insulating layer in the JJ acts as a tunneling barrier of height $V>E$, where $E$ is the kinetic energy of the Cooper pairs. Inside the barrier, the electronic wavefunction can then be parametrized as \cite{gross2005josephson}
\begin{equation}
\psi(x)=A\cosh kx+B\sinh kx,    
\label{eq:9}
\end{equation}
with $k=\sqrt{4m_e(V-E)}$ and coefficients
\begin{equation}
A=\frac{\sqrt{n_1}e^{i\varphi_1}+\sqrt{n_2}e^{i\varphi_2}}{2\cosh 2kw },~~~B=\frac{\sqrt{n_1}e^{i\varphi_1}-\sqrt{n_2}e^{i\varphi_2}}{2\sinh 2k w },   
\label{eq:10}
\end{equation}
where $m_e$ denotes the mass of the electron, $n_{1,2}$ are the Cooper pair number densities at the two superconducting electrodes, $w$ is the width of the insulator, which should be much smaller than the superconductor size $a$. Quantum tunneling of the condensate wavefunction through the classically forbidden barrier gives rise to a current. In the absence of any extra voltage and magnetic field \footnote{Electromagnetic backgrounds are suppressed in the JJ loop region by shielding such that $\mathbf{E}=\mathbf{B}=0$, and no magnetic flux is enclosed by the superconducting loop. In the simply connected configuration, a gauge can be chosen where the vector potential $\mathbf{A}=0$, eliminating ordinary EM contributions to the Josephson phase.
}, the current density, $J=(-2e/m_e)\mathrm{Re}(\psi^* i\nabla\psi)$, can be related to the phase difference as \footnote{Prior to introducing the external source, the Josephson junction is incorporated into a superconducting loop in order to null the initial phase difference.}\cite{orlando1991foundations}
\begin{equation}
J=\frac{en}{m_e w}\sin\Delta \varphi,
\label{eq:11}
\end{equation}
where we consider $n_1=n_2=n$ and $e$ denotes the electron charge.

For a single JJ, the measurable response to a small phase perturbation is a modulation of the switching current, which is bounded by the junction critical current, $I_c$. A current-biased JJ obeys
\begin{equation}
I(t)=I_c \sin\Delta\varphi(t),   ~~~I_c=\frac{\pi \Delta_s}{2eR_N}, 
\label{eq:12}
\end{equation}
where $\Delta_s\sim \mathrm{meV}$ is the superconductive energy gap and $R_N\sim \mathcal{O}(1)~\Omega$ is the normal-state junction resistance \cite{Ambegaokar:1963zz}. In realistic devices designed for precision measurements, $I_c$ typically lies in the $\mu \mathrm{A}-\mathrm{mA}$ range. For a small phase shift $\Delta\varphi\ll1$, the induced signal appears as a linear modulation of the switching current, $\delta I_s\simeq I_c\Delta \varphi$, which constitutes the experimentally accessible observable. In a SQUID (superconducting quantum interference device) operated in a flux-locked loop (FLL) \cite{DRUNG2002134}, the phase difference is equivalently mapped to an effective magnetic flux, enabling a direct and highly sensitive flux-based readout of the scalar-induced phase modulation.
In this configuration the SQUID is maintained at its optimal working point, where small variations of the junction phase are linearly transduced into a measurable output voltage or feedback flux through the SQUID transfer function.

We find that the dominant irreducible noise source arises from quantum fluctuations associated with the number-phase uncertainty of the superconducting condensate. As we will see, our projected signals exceed this quantum noise floor over a wide region of parameter space.

Additional contributions to the background include thermal noise, which induces an effective current $I_T\sim eT$.  Taking a conservative operating temperature of $T=1~\mathrm{mK}$, we obtain $I_T\sim10^{-10}~\mathrm{A}$ \cite{PhysRevLett.127.100401,Christodoulou:2023hyt}, well below the quantum-fluctuation-limited noise. Then, electromagnetic forces between macroscopic components due to quantum vacuum fluctuations generate Casimir forces; however, these geometry-dependent forces act on the superconducting surfaces and do not couple directly to the internal Cooper pair condensate, and therefore do not contribute to the accumulated phase \cite{Klimchitskaya:2015vra,Dantchev:2022hvy}. Background EM fields, including the geomagnetic field, can be efficiently suppressed using existing superconducting and magnetic shielding techniques \cite{Westphal2021}. Any residual EM fields common to both superconducting electrodes cancel in the measured phase difference.

Finally, we have assessed the role of gravitational backgrounds. The gravitational potential of the Earth can be eliminated by placing the two superconducting electrodes at the same height, ensuring cancellation in the phase difference. Moreover, the phase induced by the gravitational field of the source objects is found to be many orders of magnitude smaller than that generated by the boson-mediated interaction and can be safely neglected \cite{Cheng:2024yrn}.
 
In the following sections, we specify the form of $V_\phi(r)$ for different source configurations and interaction types, and compute the corresponding phase shifts and coupling sensitivities for different tabletop JJ-based setups.

\section{Photophilic Scalar Interactions}\label{sec1}
\begin{figure}
    \centering
    \includegraphics[width=1.0\linewidth]{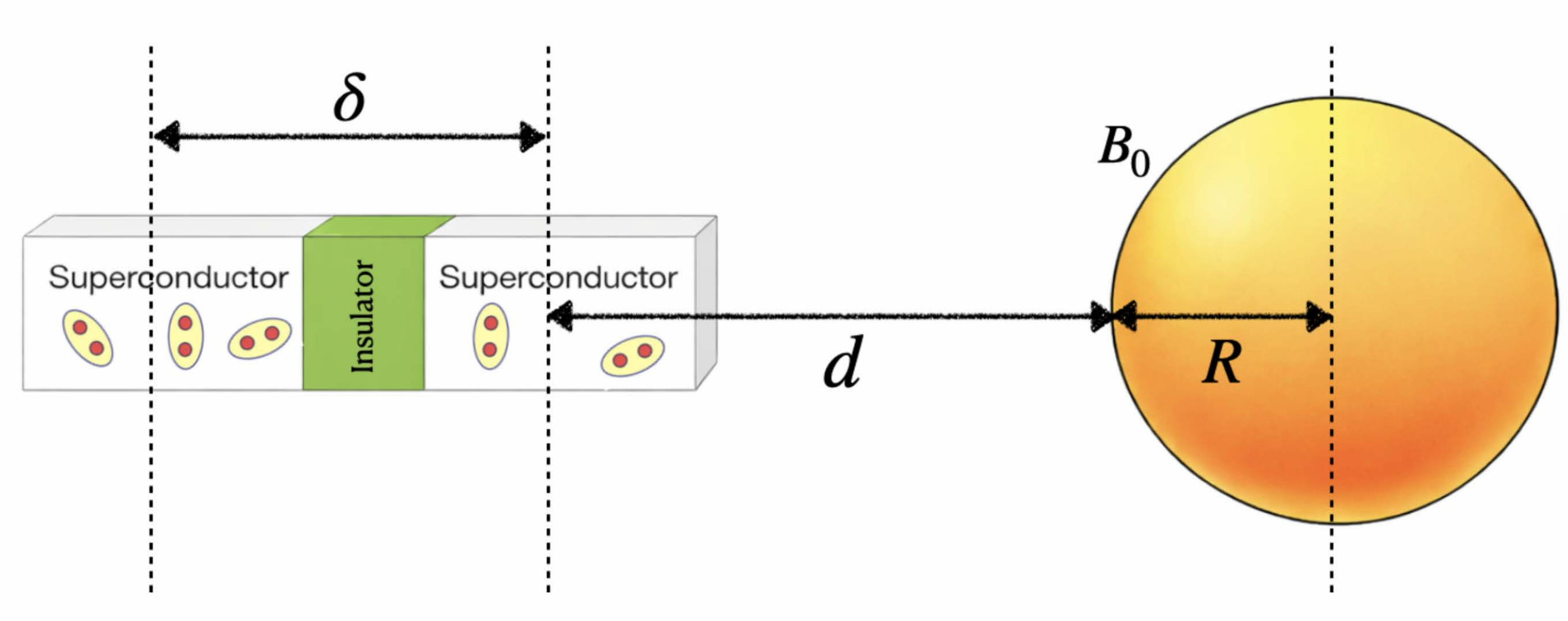}
    \caption{Schematic of a tabletop experiment featuring a magnetized sphere and an S--I--S Josephson junction (not to scale).}
    \label{one}
\end{figure}

As an instructive example, we first consider a CP-even scalar field $\phi$ of mass $m_\phi$ coupling to electromagnetism through the effective interaction \cite{Poddar:2025oew}
\begin{equation}
\mathcal{L}=\frac{1}{2}\partial_\mu \phi \partial^\mu\phi-\frac{1}{2}m^2_\phi\phi^2-\frac{1}{2}g_{\phi\gamma\gamma}\phi F_{\mu\nu}F^{\mu\nu} - g_{\phi ee} \phi \bar{e}e,
\label{eq:1}
\end{equation}
where $F_{\mu\nu}$ is the electromagnetic (EM) field-strength tensor, $g_{\phi\gamma\gamma}$ is the effective scalar-photon coupling constant, and $g_{\phi ee}$ is the scalar coupling to electrons. Varying Eq.~\eqref{eq:1} yields the equation of motion of the scalar field,
\begin{equation}
(\Box+m_\phi^2)\phi=-g_{\phi\gamma\gamma}(\mathbf{B}^2-\mathbf{E}^2) - g_{\phi ee}  \bar{e}e,  
\label{eq:l2}
\end{equation}
with $\Box$ is the d'Alembertian operator and $F_{\mu\nu}F^{\mu\nu}=2(\mathbf{B}^2-\mathbf{E}^2)$, where $\mathbf{B}$ and $\mathbf{E}$ are the magnetic and electric field strengths respectively. 

In the non-relativistic limit, $\bar e e \simeq n_e$. Since the electron density in each superconducting lead is approximately uniform, this term mainly contributes a nearly constant background to the scalar field and is subleading for the spatial variation across the junction that determines the observable phase difference. We therefore neglect it in solving for the sourced field profile.\footnote{We explicitly show the response of the superconducting condensate to the scalar-coupled Cooper-pair electron interaction in Appendix~\ref{appendix}.} For a static configuration, Eq.~\eqref{eq:l2} then reduces to a Poisson-like equation
which can be interpreted as sourcing by an effective scalar-induced charge density $\rho_\phi=g_{\phi\gamma\gamma}(\mathbf{B}^2-\mathbf{E}^2)$.

As a simple benchmark, we consider a static magnetized sphere of radius $R$ with a dipolar magnetic field outside the surface \cite{Goldreich:1969sb,Shapiro:1983du},
\begin{equation}
\mathbf{B}(r,\theta)=\frac{B_0R^3}{r^3}\Big(\cos\theta \hat{r}+\frac{\sin\theta}{2}\hat{\theta}\Big), ~~~r>R,    
\end{equation}
where $B_0$ is the surface magnetic field of the sphere and $\theta$ is the polar angle. Neglecting the effect of the electric field contribution to $\rho_\phi$ for a static configuration, the scalar field profile to the leading order in the far field region takes the Yukawa form \cite{Poddar:2025oew}
\begin{equation}
\phi(r)\simeq g_{\phi\gamma\gamma} B^2_0 R^3 \Big(\frac{e^{-m_\phi r}}{6r}\Big),   
\label{eq:2}
\end{equation}
with interaction range $\lambda\simeq m_\phi^{-1}$. Therefore, the magnetized sphere behaves as if it carries an effective scalar-induced charge $Q_\phi\simeq g_{\phi\gamma\gamma}B_0^2 R^3/6$.
We illustrate this setup in FIG.~\eqref{one}
with the hierarchy of scales $d\gg R > \delta \sim a$ such that Eq.~\eqref{eq:2} is a good approximation. 

The Cooper pair electrons feel the potential sourced by the magnetized sphere as
\begin{equation}
V_\phi(r)\simeq2g_{\phi ee }g_{\phi\gamma\gamma}B^2_0 R^3\Big(\frac{e^{-m_\phi r}}{6r} \Big),
\label{eq:3}
\end{equation} 
where the prefactor of $2$ accounts for Cooper pair electron. 
Accordingly, the phase accumulated by the superconducting Cooper pair electron in the coherent state $|1\rangle$, due to the scalar-induced potential $V_\phi$ generated by the entire magnetized sphere and evaluated at a distance $R+d$ from its center over an observation time $T$, can be expressed as
\begin{equation}
\varphi_{|1\rangle}(R+d)=\frac{1}{3}g_{\phi e e}g_{\phi\gamma\gamma}B^2_0R^3T\frac{e^{-m_\phi(R+d)}}{R+d}.    
\label{eq:5}
\end{equation}
Similarly, the phase accumulated by the superconducting Cooper pair electron in the coherent state $|2\rangle$ at a distance $(R+d+\delta)$ from the center of the magnetized sphere is given as $\varphi_{|2\rangle}(R+d+\delta)$. Thus, the phase difference between the two coherent states induced by the scalar-mediated potential becomes
\begin{eqnarray}
\Delta\varphi&=& \varphi_{|1\rangle}(R+d)-\varphi_{|2\rangle}(R+d+\delta)\nonumber\\
&\simeq&    \frac{1}{3}g_{\phi e e}g_{\phi\gamma\gamma}B^2_0R^3 T\frac{e^{-m_\phi(R+d)}}{R+d}(1-e^{-m_\phi\delta}).
\label{eq:l7}
\end{eqnarray}
In the limit $\delta<R+d$, we can write $e^{-m_\phi \delta}\simeq 1-m_\phi\delta$ and Eq.~\eqref{eq:l7} becomes
\begin{equation}
\Delta\varphi\simeq \frac{1}{3}g_{\phi e e}g_{\phi\gamma\gamma}B^2_0R^3 Tm_\phi\delta\frac{e^{-m_\phi(R+d)}}{R+d}.    
\label{eq:8l}
\end{equation}

Consequently, the accumulated phase shift grows with both the surface magnetic field strength and the radius of the sphere. In the light scalar regime, the phase shift decreases. Also, in the heavy scalar regime, the phase shift becomes exponentially suppressed. The strongest sensitivity on the mixed coupling is obtained at the scalar mass $m_\phi\sim1/(R+d)$.

\subsection{Sensitivity to photophilic scalar coupling}
\begin{figure}
    \centering
    \includegraphics[width=1.0\linewidth]{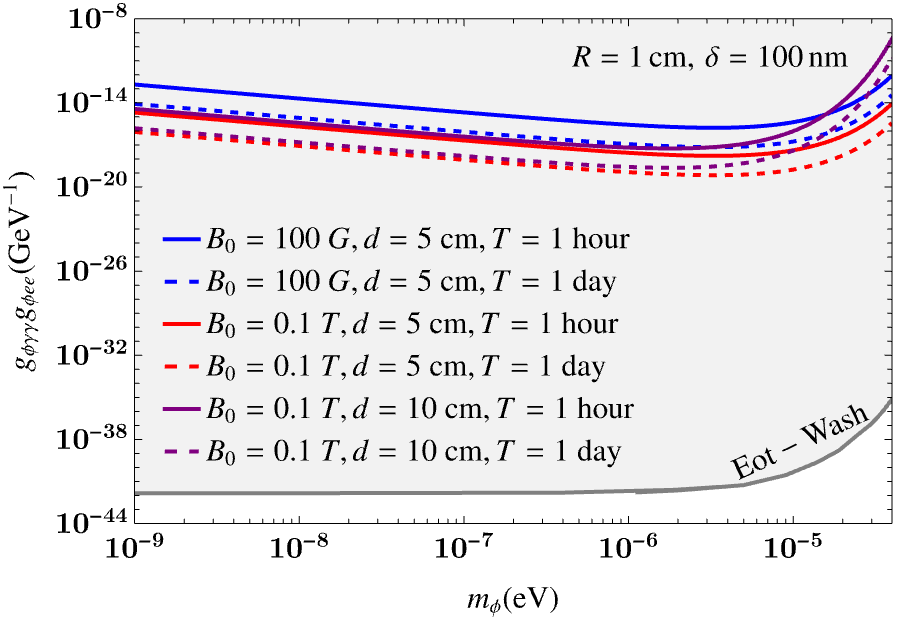}
    \caption{Projected sensitivities to the electron-scalar-photon mixed coupling $g_{\phi ee} g_{\phi\gamma\gamma}$ for scalar masses in the range $10^{-9}~\mathrm{eV} \lesssim m_\phi \lesssim 10^{-4}~\mathrm{eV}$. The sensitivities are shown for different choices of surface magnetic field values, JJ-slab distance, and operational timescale of the JJ-sphere tabletop setup, as indicated by the colored lines. The gray shaded region correspond to benchmark bounds obtained by multiplying existing constraints on the individual couplings $g_{\phi ee}$ and $g_{\phi\gamma\gamma}$ from Eot-Wash experiment.
   }
    \label{fig:two}
\end{figure}

Within the semiclassical London framework, the Cooper pair density can be written as \cite{London1935}
\begin{equation}
n=\frac{m_e}{e^2\lambda^2_L},
\label{eq:8}
\end{equation}
where $\lambda_L$ is the London penetration depth. To estimate the sensitivity to the mixed coupling $g_{\phi ee} g_{\phi\gamma\gamma}$, we consider a superconducting junction with linear size $a = 100~\mathrm{nm}$ and London penetration depth $\lambda_L = 50~\mathrm{nm}$. In this limit, the Cooper-pair number densities in the two superconducting states are identical, $n_1 = n_2 \simeq 1.2 \times 10^{22}~\mathrm{cm}^{-3}$ (using Eq.~\eqref{eq:8}), yielding a total number of Cooper pairs $N \simeq 1.2 \times 10^{7}$ within the superconducting volume.

Since a superconducting condensate is described by a coherent state, the particle number and phase uncertainties obey the number-phase uncertainty relation $\Delta N\, \Delta\varphi \gtrsim 1$ \cite{PhysRevLett.14.387}. Taking the intrinsic number fluctuation to be $\Delta N \simeq \sqrt{N}$, the corresponding phase sensitivity reaches $\Delta\varphi_c \simeq 2.8 \times 10^{-4}$. Consequently, any scalar-induced Josephson phase shift satisfying $\Delta\varphi > \Delta\varphi_c$ can be resolved beyond the quantum noise limit, which is expected to be the primary noise source of this setup.

For the external source, we consider a magnetized sphere with surface magnetic field $B_0 = 100~\mathrm{G}$ and radius $R = 1~\mathrm{cm} $. The separation between the two superconducting leads is taken to be $\delta = 100~\mathrm{nm}$, while the distance from the sphere's surface to the center of the nearest superconductor is $d = 5~\mathrm{cm}$, satisfying the hierarchy $a \sim \delta < R\ll d $. We further assume a total experimental running time of $T = 1~\mathrm{hour}(1~\mathrm{day})$. For a representative scalar-induced phase shift $\Delta\varphi = 3 \times 10^{-4}$, corresponding to a Josephson current $I \simeq 1.5~\mathrm{\mu A}$ for $I_c\sim 5~\mathrm{mA}$ with proper choice of S--I--S JJ (c.f. Eq.~\eqref{eq:12}), we derive the projected sensitivity to the mixed coupling $g_{\phi ee} g_{\phi\gamma\gamma}$, shown by the blue solid (dashed) curve in FIG.~\eqref{fig:two}. We also present the corresponding sensitivity assuming an enhanced magnetic field $B_0 = 0.1~\mathrm{T}$ and an operational timescale $T = 1~\mathrm{hour}(1~\mathrm{day})$, shown by the red solid (dashed) curve, with all other parameters kept fixed.

In addition, we consider a configuration with increased separation distances, $d = 10~\mathrm{cm}$, while maintaining other parameters fixed except the operational timescale. We obtain the corresponding sensitivities shown by the purple solid and dashed curves for $(B_0, T) = (0.1~\mathrm{T}, 1~\mathrm{hour})$ and $(0.1~\mathrm{T}, 1~\mathrm{day})$, respectively. Increasing the value of $d$ relatively shifts the minimum of the curve towards left of the scalar-mass axis (relatively long-ranged), as expected.

This experimental strategy yields direct sensitivities to the mixed coupling $g_{\phi ee}\,g_{\phi\gamma\gamma}$ at the centimeter scale using a single tabletop experimental configuration.
We compare our projected sensitivity curves with benchmark estimates obtained by taking the simple product of existing individual bounds on $g_{\phi\gamma\gamma}$ and $g_{\phi ee}$, derived from Eot-Wash experiment \cite{Hees:2018fpg}, which yields the strongest constraints on the product of individual couplings for the parameter space of interest. 

The resulting bound is weaker than the existing limits on the simple product coupling. The primary reason is that the scalar-induced effective EM charge of the magnetized sphere is extremely small, given as
\begin{equation}
Q_\phi\simeq 8\times 10^{-14}\Big(\frac{g_{\phi\gamma\gamma}}{10^{-20}~\mathrm{GeV^{-1}}}\Big)\Big(\frac{B_0}{0.1~\mathrm{T}}\Big)^2\Big(\frac{R}{1~\mathrm{cm}}\Big)^3,
\end{equation}
as obtained from Eq.~\eqref{eq:2}. Consequently, the induced charge is negligibly small compared with the ordinary baryonic charge scale of a macroscopically neutral sphere.
However, we emphasize that such products do not rigorously correspond to constraints on the mixed coupling $g_{\phi e e} g_{\phi\gamma\gamma}$.
Combining them assumes that both bounds apply simultaneously within the same physical environment and parameter regime, an assumption that can break down in scenarios with environment-dependent scalar couplings \cite{OHare:2020wah,Poddar:2023bgk}.

Nevertheless, in the absence of dedicated experimental constraints on the mixed coupling $g_{\phi e e} g_{\phi\gamma\gamma}$, we adopt this procedure as a conservative benchmark and include these product bounds for comparison with our projected sensitivities. 
 
\subsection{Readout for a JJ-magnetized sphere in probing mixed coupling $g_{\phi e e}g_{\phi\gamma\gamma}$}

To isolate the scalar-mediated signal from slow drifts and instrumental backgrounds, the measurement can be performed using controlled modulation schemes. A convenient approach is to modulate the distance between the magnetized sphere and the junction around a mean value $d_0$, 
\begin{equation}
d(t)=d_0+\Delta d\cos(2\pi f_d t),
\label{readout2}
\end{equation}
with $\Delta d\ll d_0$ and $f_d$ the modulation frequency. Since the scalar-mediated potential depends on the separation, the induced current response can be expanded as
\begin{equation}
\Delta I(t)\simeq \Delta I(d_0) +\frac{\partial \Delta I}{\partial d}\Bigg|_{d_0} \Delta d   \cos(2\pi f_d t)+\cdots.
\label{readout3}
\end{equation}
The scalar signal therefore appears at the known modulation frequency $f_d$, enabling extraction via lock-in detection. This technique suppresses dc offsets and slow drifts while isolating the characteristic distance dependence of the scalar interaction.

An additional discriminator can be obtained by modulating the magnetization of the sphere. If the remnant surface magnetic field is varied around a mean value $B_0$ as
\begin{equation}
B_0(t)=B_0+\Delta B\cos(2\pi f_B t),
\end{equation}
the scalar-induced signal follows the corresponding change in the magnetic energy density that sources the scalar field. The response can therefore be detected at the modulation frequency $f_B$, providing a complementary lock-in channel. This scaling with the magnetic energy density distinguishes the scalar signal from many electromagnetic spurious effects such as flux pickup generated by mechanical motion.

Systematic effects can be further quantified through geometric null configurations. For instance, replacing the magnetized sphere with a non-magnetized sphere of identical mass and geometry allows one to isolate purely mechanical and electrostatic couplings. Repeating the measurement for several values of the mean separation $d_0$ also tests the predicted short-range attenuation of the scalar-mediated potential. In this way the SQUID-based JJ acts as a phase sensitive probe of the photophilic scalar interaction.

A crucial requirement to perform the experiment is that the stray magnetic flux generated by the magnetized sphere does not penetrate the JJ readout loop. For the benchmark geometry ($R\sim1~\mathrm{cm}$, $d\sim 5~\mathrm{cm}$, $B_0\sim 0.1~\mathrm{T}$), an unconfined dipole source would produce a stray field at the junction of order $B(R+d)\approx B_0(R/(R+d))^3\sim \mathcal{O}(5\times 10^{-4}~\mathrm{T})$. In practice, the source magnetization is therefore made magnetically self-contained with compensation coils so that only a small leakage field reaches the SQUID loop \cite{Yakhmi2021}. The JJ/SQUID is then operated inside multilayer high-permeability and superconducting shielding and in a flux-locked loop configuration, which nulls the enclosed magnetic flux rather than the local field. To maintain stable operation and avoid flux-induced phase drift, the residual flux must satisfy $\Phi_{\mathrm{leak}}\ll \Phi_0$; a typical criterion is $\Phi_{\mathrm{leak}}\lesssim 10^{-3} ~\Phi_0$, where $\Phi_0=2\pi/2e=2.07\times 10^{-15}~\mathrm{Wb}$, implies residual fields $B_{\mathrm{leak}}\lesssim10\textit{-}100~\mathrm{nT}$ for a typical loop area $A_{\mathrm{eff}}\sim (10~\mathrm{\mu m})^2$, achievable with standard SQUID practice and zero-field cooldown \cite{clarke2006squid,Fagaly2006SuperconductingQI}. Magnetic shielding suppresses stray EM fields but does not affect the scalar field sourced by the magnetic energy density inside the magnetized sphere. Ordinary magnetic pickup scales linearly with the residual field ($\propto B_{\mathrm{leak}}$), whereas the photophilic scalar signal scales with the magnetic energy density $(\propto B^2_0)$, providing a discriminator against the leakage background. 

\section{Lorentz-violating scalar interactions}\label{sec2}
\begin{figure}
    \centering
    \includegraphics[width=1.0\linewidth]{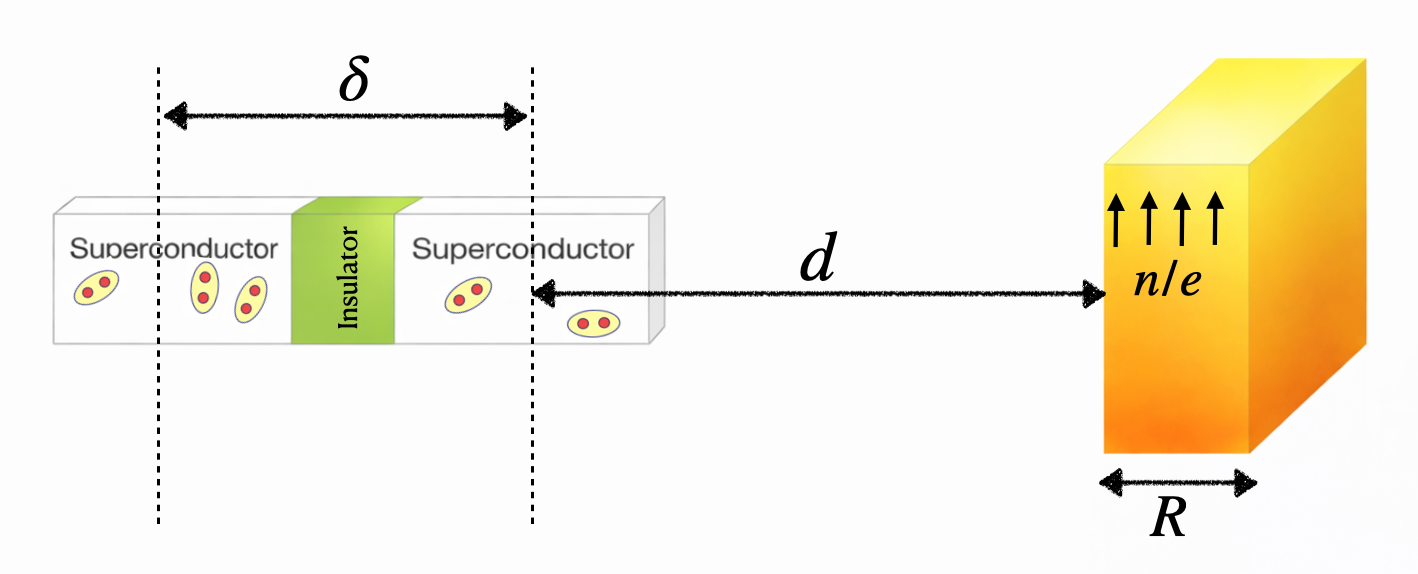}
    \caption{Schematic of a tabletop experiment featuring a  plate of polarized electrons/neutrons and an S--I--S Josephson junction (not to scale).}
    \label{three}
\end{figure}

In this section we consider an alternative configuration in which a slab of thickness $R$, composed of polarized fermions (electrons or neutrons), is placed in the vicinity of an S--I--S JJ as shown schematically in FIG.~\eqref{three}. The goal is to probe a Josephson current induced by a Lorentz-violating scalar-mediated potential, in which the scalar field couples to the polarized fermions in the slab through a Lorentz-violating vertex of the corresponding Feynman diagram and, at another vertex, to the coherent Cooper-pair condensate in the superconductor. As in the previous setup, the slab is positioned at a distance $d$ from the midpoint of one superconducting lead, while the separation between the two coherent states is $\delta$, with the hierarchy of scales $R>d\gtrsim \delta\sim a$. This configuration enables the probing of new bosonic interactions at micrometer length scales, in contrast to the previous setup, which was sensitive primarily to interactions at the centimeter scale.

The relevant Lorentz-violating scalar-mediated interaction Lagrangian can be written as \cite{Altschul:2012xu,Carenza:2025jwn}
\begin{equation}
\mathcal{L}=\frac{1}{2}\partial_\mu \phi \partial^\mu \phi-\frac{1}{2}m^2_\phi \phi^2-\phi\bar{\psi}\Big(J_\mu\gamma_5 \gamma^\mu+\frac{1}{2}L_{\mu\nu}\sigma^{\mu\nu}\Big)\psi, 
\label{2a}
\end{equation}
where $\psi$ denotes the polarized fermions in the slab, $J_\mu$ is a dimensionless Lorentz-violating pseudovector coefficient and $L_{\mu\nu}$ is a dimensionless Lorentz-violating antisymmetric tensor. In addition, $J_\mu$ violates CPT (charge conjugation-parity-time reversal) symmetry, since the associated operator is odd under CPT. Furthermore, the scalar field interacts linearly with the Cooper pair electron number density, which is CPT and Lorentz-conserving.

Assuming that the scalar boson of mass $m_\phi$ mediates interaction between the polarized fermions in the source slab and unpolarized spin singlet state Cooper pair electrons in the superconducting leads, the spatial component of $J_\mu^e$ give rise to a spin-dependent potential in the non-relativistic limit of fermions as \cite{Altschul:2012xu,Carenza:2025jwn}
\begin{equation}
V_\phi(r)=(\bm{\sigma}_\psi\cdot \tilde{\bm{J}}_\psi)\frac{2 g_{\phi e e} e^{-m_\phi r}}{4\pi r},   
\label{2b}
\end{equation}
where $\bm{\sigma}_\psi$ denotes the fermion spins in the slab, $\tilde{J}_j = J_j + \epsilon_{jkl}L_{kl}$, and $g_{\phi e e}$ denotes the scalar-electron coupling in the superconducting condensate. The range of this new spin-dependent Lorentz-violating scalar-mediated potential is given as $\lambda=1/m_\phi$. The temporal component of $J^e_\mu$ leads to a velocity-dependent potential which we do not consider here.

Given that the detector sample is much smaller than the source slab, the superconducting lead can be approximated as a point-like object. As a result, the field within the detector is treated as uniform, and the source, from the perspective of the detector, effectively appears as an infinite plane.
Therefore, integrating all possible fermion coordinates of the slab, the phase accumulated by the Cooper pair electron in the superconducting coherent state $|1\rangle$ due to this scalar-mediated potential over a running time $T$ is given by
\begin{equation}
\varphi_{|1\rangle}(d)= 2g_{\phi e e} \bm{\tilde{J}}_\psi n_pT\int^R_0 \int^\infty_0 \frac{e^{-m_\phi\sqrt{r^2+(d+z)^2}}}{4\pi \sqrt{r^2+(d+z)^2}} 2\pi r dr dz,  
\label{3b}
\end{equation}
where we consider $\bm{\tilde{J}}_\psi$ is along the direction of polarized spins, which is assumed to be radial \footnote{The spins may alternatively be aligned along the $\hat{z}$ direction, which introduces an angular projection factor of $\tilde{\bm{J}}$ along $\hat{z}$ in the integrand. We demonstrate this configuration explicitly in the axion-mediated monopole-dipole case, discussed below.}. The number density of polarized fermions in the slab is denoted as $n_p$, and the height of the slab is taken to be sufficiently large to be approximated as infinite.
Evaluating the integral in Eq.~\eqref{3b} yields
\begin{equation}
\varphi_{|1\rangle}(d)= g_{\phi e e} \bm{\tilde{J}}_\psi n_pT \frac{e^{-m_\phi d}(1-e^{-m_\phi R})}{m^2_\phi}.  
\label{4b}
\end{equation}

Analogously, the phase accumulated by the Cooper pair electron in the second superconducting coherent state is $\varphi_{|2\rangle}(d+\delta)$. The resulting phase difference between the two states is therefore
\begin{eqnarray}
\Delta \varphi&=&\varphi_{|1\rangle}(d)- \varphi_{|2\rangle}(d+\delta)\nonumber\\
&\simeq &\frac{g_{\phi e e} \bm{\tilde{J}}_\psi n_pT}{m^2_\phi}e^{-m_\phi d}(1-e^{-m_\phi \delta})(1-e^{-m_\phi R}).
\label{5b}
\end{eqnarray}
The corresponding Josephson current induced by the scalar-mediated phase shift is then obtained as
\begin{equation}
I\simeq I_c\frac{g_{\phi e e} \bm{\tilde{J}}_\psi n_pT \delta}{m_\phi}e^{-m_\phi d},
\label{6b}
\end{equation}
where we have expanded to leading order in the separation $\delta$ and take $e^{-m_\phi R}\sim 0$ because of the large thickness of the slab.

\subsection{Sensitivity to Lorentz-violating coupling}
\begin{figure}
    \centering
    \includegraphics[width=1.0\linewidth]{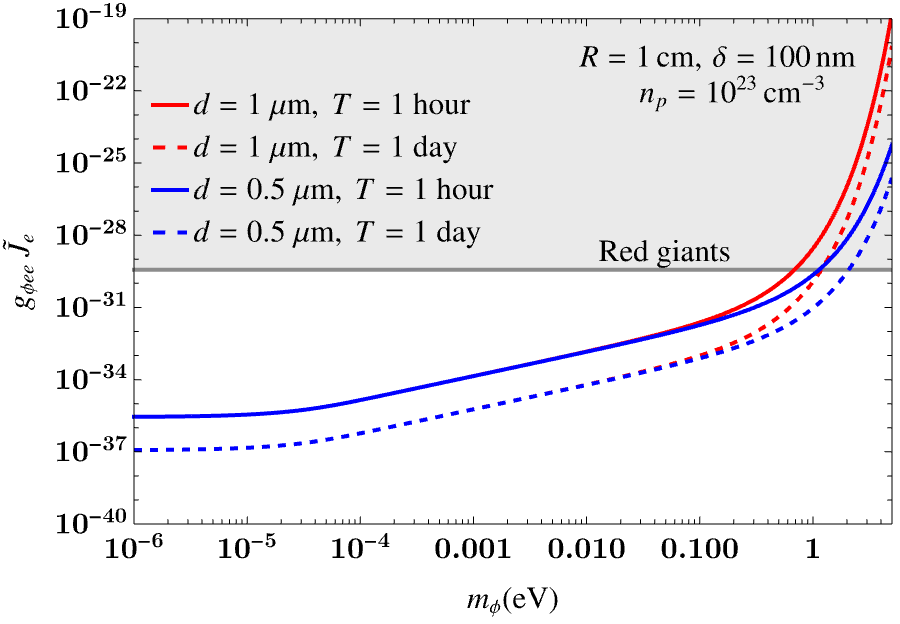}
    \caption{Projected sensitivity to the scalar-mediated Lorentz-violating mixed coupling $g_{\phi e e}\tilde{J}_e$ for scalar masses in the range $10^{-6}~\mathrm{eV}\lesssim m_\phi\lesssim 5~\mathrm{eV}$. The sensitivities are shown for different choices of JJ-slab distance $(d)$ and running time $(T)$ of the tabletop setup, as indicated by the colored lines. The gray shaded regions correspond to benchmark bounds obtained by multiplying existing constraints on the individual couplings $g_{\phi e e}$ and $\tilde{J}_e$ from red giants.}
    \label{kfour}
\end{figure}
In estimating the sensitivity to Lorentz-violating couplings, we consider a superconducting JJ characterized by $\lambda_L = 50~\mathrm{nm}$ and $a = 100~\mathrm{nm}$. For these parameters, the Cooper pair number densities in the two coherent states are equal, $n_1 = n_2 \simeq 1.2 \times 10^{22}~\mathrm{cm^{-3}}$, corresponding to a total number of Cooper pairs $N \simeq 1.2 \times 10^{7}$. The resulting phase sensitivity, limited by quantum fluctuations, is therefore $\Delta\varphi_c \sim 1/\sqrt{N} \simeq 2.8 \times 10^{-4}$. Consequently, any Lorentz-violating scalar-induced Josephson phase shift satisfying $\Delta\varphi > \Delta\varphi_c$ is, in principle, detectable beyond the quantum noise limit.

For the source configuration, we consider an electron spin-polarized slab of thickness $R = 1~\mathrm{cm}$ placed at a distance $d = 1~\mathrm{\mu m}$ from the superconducting element, with $\delta \simeq a \simeq 100~\mathrm{nm}$, satisfying the geometric hierarchy $a \sim \delta \lesssim d < R$. The number density of polarized spins in the slab is taken to be $n_p = 10^{23}~\mathrm{cm^{-3}}$, which is achievable with present technologies \cite{Heckel:2008hw}.
Assuming an operational integration time of $T = 1~\mathrm{hour}$ ($1~\mathrm{day}$), we obtain strong projected sensitivities to the Lorentz-violating scalar-mediated mixed coupling $g_{\phi e e}\tilde{J}_e$, shown by the red solid (dashed) curves in FIG.~\eqref{kfour}. We also present sensitivities for a reduced JJ-slab separation $d = 0.5~\mathrm{\mu m}$, indicated by the blue solid (dashed) curves, for the same choices of integration time. We use $\Delta\varphi\simeq 3\times 10^{-4}$ to obtain the sensitivities on the mixed coupling.

Constraints on the scalar-mediated Lorentz-violating mixed coupling have been derived only recently \cite{Carenza:2025jwn}. However, existing limits predominantly probe very light scalars, $m_\phi\lesssim \mathcal{O}(10^{-10})~\mathrm{eV}$. At higher masses, around $m_\phi \sim 0.01~\mathrm{eV}$, the only available constraint originates from stellar cooling arguments in red giants by multiplying individual couplings. Such astrophysical bounds are subject to potentially significant systematic uncertainties associated with stellar modeling and background effects.

Currently, there are no direct experimental or observational bounds on this mixed coupling $g_{\phi e e}\tilde{J}_e$ in the mass range $10^{-6}~\mathrm{eV} \lesssim m_\phi \lesssim 5~\mathrm{eV}$. For comparison, we construct a benchmark constraint by multiplying the only existing individual limits on $g_{\phi e e}$ and $\tilde{J}_e$ in this scalar mass range derived from red giant cooling arguments \cite{Carenza:2025jwn}, and overlay it with our projected sensitivity. In contrast to the mediator scenario considered here, the red giant constraints on the individual couplings are derived by treating the scalar as an emitted final state particle from either the scalar interaction vertex or the Lorentz-violating vertex. For the benchmark parameter choices considered here, our setup improves upon the red giant limits by approximately seven orders of magnitude for $m_\phi\simeq 10^{-6}~\mathrm{eV}$ and about two orders of magnitude stronger for $m_\phi\simeq 1~\mathrm{eV}$.

The projected sensitivities correspond to resolvable Josephson phase shifts satisfying $\Delta\varphi \simeq 3 \times 10^{-4}$, which translates to a measurable Josephson current of order $1.5~\mathrm{\mu A}$, for $I_c\simeq~5~\mathrm{mA}$.
Moreover, the projected sensitivities 
are obtained by assuming maximal projection, i.e. by taking the Lorentz-violating vector $\tilde{\boldsymbol J}_e$ to be aligned with the polarization direction of the spin-polarized source. Under this assumption, the spin-vector contraction is absorbed into an effective coupling $\tilde{\boldsymbol J}_e$. If instead the Lorentz-violating vector $\tilde{\boldsymbol J}_e$ is not aligned with the spin-polarization axis $\hat{\boldsymbol\sigma}_e$, the projection $\boldsymbol\sigma_e\cdot\tilde{\boldsymbol J}_e$ varies in the laboratory frame due to the Earth's rotation. This leads to a sidereal modulation of the induced spin precession. Importantly, the directional dependence in this case is referenced to a fixed cosmic vector $\tilde{\boldsymbol J}_e$, rather than to the geometric orientation of the source. As a consequence, the projection $\boldsymbol\sigma_e\cdot\tilde{\boldsymbol J}_e$ oscillates with a period of one sidereal day.

A readout scheme for probing the Lorentz-violating scalar-mediated interactions is presented below. 

\subsection{Readout for a JJ-polarized slab in probing $g_{\phi e e}\tilde{\boldsymbol J}_e$}

We now outline a measurement strategy to detect the Lorentz-violating scalar-mediated interaction in the polarized slab-JJ setup shown in FIG.~\eqref{three}. The observable is a scalar-induced phase difference between the two superconducting coherent states, which manifests as a modulation of the Josephson current. Equivalently, this phase shift can be measured as a flux signal in a SQUID readout. In practice, the junction can be embedded in a SQUID operated in a FLL, where the phase perturbation $\Delta\varphi$ appears as an equivalent flux variation $\Delta\Phi$. The SQUID output is therefore directly proportional to $\Delta\Phi(t)$ through the SQUID transfer function.

The signal in Eq.~\eqref{5b} is proportional to the projection $\tilde{J}_{e\parallel}=\mathbf{\tilde J}_e\cdot \hat{\bm \sigma}_e$. To isolate this contribution and suppress systematic effects, we employ several modulation strategies. First, the slab polarization is periodically reversed, $\hat{\bm \sigma}_e\rightarrow-\hat{\bm \sigma}_e$, which flips the sign of the induced current and the phase shift. Demodulation at the chosen modulation frequency $f_{\rm mod}$, selected within a band where the readout noise is approximately white, provides an estimate of the current variation $\widehat{\Delta I}$ that is robust against slow drifts and static offsets. Second, measurements can be performed for different slab-junction separations $d_i$. A genuine scalar-mediated interaction obeys the Yukawa scaling $\widehat{\Delta I}(d)\propto e^{-m_\phi d}$, allowing an internal consistency check and a mass-dependent analysis. Finally, because the preferred-frame vector $\mathbf{\tilde J}_e$ is fixed in space, the projection $\tilde{J}_{e\parallel}(t)$ modulates with the rotation of the Earth. After demodulation at $f_{\rm mod}$, the extracted signal amplitude can be fitted as

\begin{equation}
\widehat{\Delta I}(t)=\Delta I_0+\Delta I_1\cos(\omega_\oplus t)+\Delta I_2\sin(\omega_\oplus t),
\end{equation}

where $\omega_\oplus$ is the sidereal frequency. The coefficients $\Delta I_0$, $\Delta I_1$, and $\Delta I_2$ probe the components of $\mathbf{\tilde J}_e$ along and transverse to Earth's rotation axis, providing a distinctive signature that separates Lorentz-violating effects from terrestrial systematics. The sidereal modulation readout is similar to what is described in \cite{Heckel:2008hw}.

\section{Axion-mediated monopole-dipole interaction}\label{sec3}

Considering the same experimental configuration illustrated in FIG.~\eqref{three}, a monopole-dipole interaction mediated by an axion $a$ can also be probed through Josephson current measurements. In this case, we assume that the axion couples to polarized fermions (electrons or nucleons) with a pseudoscalar interaction in the slab at one vertex, while it couples to the Cooper pair electrons with a scalar interaction in the superconducting condensate at the other vertex. The relevant axion-mediated monopole-dipole interaction Lagrangian can be written as \cite{Moody:1984ba,Daido:2017hsl}
\begin{equation}
\mathcal{L}=\frac{1}{2}\partial_\mu a \partial^\mu a-\frac{1}{2}m^2_a a^2-2g_S a \bar{e}e-ig_P a \bar{\psi}\gamma_5\psi,
\label{1c}
\end{equation}
which gives rise to a monopole (scalar)-dipole (pseudoscalar) potential in the non-relativistic limit of the fermions as \cite{Moody:1984ba,Daido:2017hsl}
\begin{equation}
V(r)=\frac{2g_P g_S}{4\pi m_\psi}(\bm \sigma_\psi\cdot \hat{\bm r})\Big(\frac{m_a}{r}+\frac{1}{r^2}\Big) e^{-m_a r },
\label{2c}
\end{equation}
where $m_\psi$ denotes the mass of the polarized fermion and $\sigma_\psi$ denotes its spin. This interaction induces a macroscopic force that violates both parity and time-reversal symmetry. The scalar interaction of the axion explicitly violates CP symmetry. In the presence of CP violation, the axion can therefore acquire both scalar and pseudoscalar couplings \cite{Chang:1985mu,Barbieri:1981rs}. Such scalar interactions shift the minimum of the axion potential away from its strong-CP-preserving value, corresponding to $\theta_{\rm eff}=0$.

Consider that the fermion spins of the slab are polarized along the $\hat{\bm z}$ direction and hence 
\begin{equation}
\bm \sigma\cdot \hat{\bm r}=\cos\theta =\frac{d+z}{\sqrt{r^2+(d+z)^2}}.    
\end{equation}

We calculate the phase accumulated by the Cooper pair electron in superconducting coherent state $|1\rangle$ due to this axion-mediated monopole-dipole potential sourced by the polarized fermion slab over an observation time $T$ as
\begin{widetext}
\begin{equation}
\begin{split}
\varphi_{|1\rangle}(d)=\frac{2g_Pg_S n_p T}{4\pi m_\psi}\int ^R_0 \int ^\infty_0 e^{-m_a \sqrt{r^2+(d+z)^2}} \frac{d+z}{\sqrt{r^2+(d+z)^2}}\Bigg(\frac{m_a}{\sqrt{r^2+(d+z)^2}}+
\frac{1}{r^2+(d+z)^2} \Bigg)
2\pi r dr dz,  
\label{3c}
\end{split}
\end{equation}
\end{widetext}
where $n_p$ denotes the number density of the polarized fermions in the slab and we integrate over all possible fermion coordinates of the slab. Hence, solving Eq.~\eqref{3c} we obtain 
\begin{equation}
 \varphi_{|1\rangle}(d)=\frac{g_P g_S n_p T}{m_\psi m_a}e^{-m_a d}(1-e^{-m_a R}),  
 \label{3d}
\end{equation}
and the resulting phase difference between the two superconducting leads is
\begin{eqnarray}
\Delta \varphi&=&\varphi_{|1\rangle}(d)- \varphi_{|2\rangle}(d+\delta)\nonumber\\
&\simeq& \frac{g_P g_S n_p T}{m_\psi m_a}e^{-m_a d}(1-e^{-m_a\delta})(1-e^{-m_a R}). 
\end{eqnarray}
The corresponding Josephson current induced by this axion-mediated phase shift is then given by 
\begin{equation}
I\simeq I_c\frac{g_P g_S n_p T}{m_\psi m_a}e^{-m_a d}(1-e^{-m_a\delta})(1-e^{-m_a R}).
\end{equation}
Note, in the limit $m_a\delta\ll1$, there would be no $1/m_a$ enhancement.

\subsection{Sensitivity to axionic monopole-dipole coupling}
\begin{figure}
    \centering
    \includegraphics[width=1.0\linewidth]{mdf2.png}
    \caption{Projected sensitivity to the axion-mediated monopole-dipole coupling $g_{\rm S e}g_{\rm P e}$ for axion masses in the range $10^{-4}~\mathrm{eV}\lesssim m_a\lesssim 5~\mathrm{eV}$. The sensitivities are shown for different choices of the input parameters of the JJ-slab tabletop setup, as indicated by the colored lines. The gray shaded regions correspond to benchmark bounds obtained by multiplying existing constraints on the individual couplings $g_{P e}$ from red giants in the $\omega$ Centauri and scalar-electron coupling from fifth force and white dwarfs.}
    \label{four}
\end{figure}
In estimating the sensitivity to the axion-mediated monopole-dipole coupling, we adopt the same superconducting parameters as in the Lorentz-violating case, namely a London penetration depth $\lambda_L = 50~\mathrm{nm}$ and a transverse dimension $a = 100~\mathrm{nm}$. For this configuration, the phase sensitivity limited by quantum fluctuations is $\Delta\varphi_c \simeq 2.8 \times 10^{-4}$. Consequently, any axion-induced Josephson phase shift satisfying $\Delta\varphi > \Delta\varphi_c$ is, in principle, resolvable beyond the quantum noise limit, which is the dominant noise source of this tabletop setup.

For the source configuration, we consider the same spin-polarized electron slab as in the previous section, with thickness $R = 1~\mathrm{cm}$, positioned at a distance $d = 1~\mathrm{\mu m}$ from the superconducting element. The separation between the two coherent states is taken to be $\delta \simeq a \simeq 100~\mathrm{nm}$, satisfying the geometric hierarchy $a \sim \delta \lesssim d < R$. The number density of polarized electron spins in the slab is fixed at $n_p = 10^{23}~\mathrm{cm^{-3}}$.

The projected sensitivities to the coupling are shown by the colored curves and are evaluated for a resolvable Josephson phase shift of $\Delta\varphi = 3 \times 10^{-4}$. The red solid (dashed) curves correspond to a source-detector separation of $d = 1~\mathrm{\mu m}$ and an operational integration time of $T = 1~\mathrm{yr}~(10~\mathrm{yrs})$, while the blue solid (dashed) curves denote the case $d = 0.5~\mathrm{\mu m}$ for the same respective running times. We also show the sensitivity on the mixed coupling for $d=0.5~\mathrm{\mu m}$ with $T=1~\mathrm{hour}~(1~\mathrm{day})$ by purple solid (dashed) line. We emphasize that the sensitivity in this case should not be considered as weaker, since the existing limits are obtained by combining individual bounds on the couplings in the electronic sector, derived from different observations. These existing bounds correspond to different Feynman diagrams therefore do not directly correspond to the process considered in the present analysis, which has considered single experimental setup and a single Feynman diagram with different couplings at the two vertices. The sensitivities are obtained over the axion mass range $10^{-4}~\mathrm{eV} \lesssim m_a \lesssim 5~\mathrm{eV}$ and are compared with the product of existing constraints on the individual couplings $g_{Se}$ and $g_{Pe}$.

Dedicated laboratory searches for the axion-mediated monopole-dipole interaction have been performed only in the low-mass regime, probing axion masses $m_\phi\lesssim \mathcal{O}(10^{-4})~\mathrm{eV}$ within a single experimental framework \cite{Heckel:2008hw,OHare:2020wah}. In contrast, for heavier axions, $m_\phi\gtrsim \mathcal{O}(10^{-2})~\mathrm{eV}$, no direct laboratory constraints currently exist \cite{AxionLimits}. The available limits in this mass range are instead inferred indirectly by combining independent bounds on the individual couplings from astrophysical observations \cite{Hardy:2016kme,Capozzi:2020cbu}.

In this mass interval, the most stringent constraint on the pseudoscalar electron coupling $g_{Pe}$ arises from red giant cooling observations in $\omega$ Centauri. To construct a benchmark for comparison, this bound is combined with existing limits on the scalar-electron coupling derived from fifth-force experiments and white dwarf observations. While dedicated laboratory searches have placed constraints on mixed couplings such as $g_{Sn} g_{Pe}$, there are currently no direct experimental limits on the electron-electron mixed coupling $g_{Se} g_{Pe}$, beyond those inferred by multiplying individual bounds in the mass range of our interest.

Relative to this benchmark, the projected sensitivities presented here improve upon existing limits by up to one order of magnitude for axion masses in the range $\mathcal{O}(0.01\text{–}1~\mathrm{eV})$, with the particular choice of the experimental parameters.

\subsection{Readout for a JJ-polarized slab in probing $g_{Se}g_{Pe}$}
Instrumentally, the readout, modulation scheme, and noise properties are identical for the axion-mediated monopole-dipole interaction and for the Lorentz-violating scalar scenario. In both cases, the signal is extracted as a spin-dependent Josephson phase shift (or equivalently a current modulation) using polarization reversal and lock-in detection. The key distinction lies in the temporal structure of the extracted signal.

For the monopole-dipole axion interaction, after demodulation at $f_{\rm mod}$ the signal is expected to be time independent, apart from controlled variations associated with changes in $d$ or $\delta$. The readout strategy is very similar to the QUAX-gpgs experiment \cite{Crescini:2016lwj}. In contrast, for the Lorentz-violating scalar interaction, the extracted amplitude generally exhibits sidereal modulation, since it depends on the projection of a preferred-frame coefficient onto the laboratory polarization axis. Fitting the demodulated signal to sidereal harmonics therefore provides an additional discriminator characteristic of Lorentz-violation searches.

\section{Conclusions and discussions}\label{sec4}

We obtain sensitivities on the mixed coupling of ultralight boson-mediated interactions in the mass range $10^{-6}~\mathrm{eV}\lesssim m_{\phi,a} \lesssim 5~\mathrm{eV}$ using JJ interferometry. The proposed setups probe interaction range as small as $d\sim\mathcal{O}(0.5~\mathrm{\mu m})$, a regime that is challenging for existing tabletop experiments such as torsion-balance measurements \cite{Adelberger:2009zz}.

The first example we described is of a  JJ-magnetized-sphere configuration with which we obtain sensitivity to the mixed electron-scalar-photon coupling $g_{\phi ee}g_{\phi\gamma\gamma}$. 
A complementary JJ-electron-spin-polarized slab setup allows us to probe Lorentz-violating scalar interactions through the mixed coupling $g_{\phi ee}\tilde{J}_e$. In addition, the same experimental strategy can be employed to search for axion-mediated monopole-dipole interactions characterized by the coupling product $g_{Se}g_{Pe}$. Importantly, the sensitivities derived here do not rely on multiplying independent constraints on individual couplings, but instead arise from configurations in which distinct physical sources appear at the two vertices of the boson-mediated interaction. For comparison, we also benchmark our results against the product of existing individual bounds and find that, even in this conservative approach, the proposed JJ-based setups remain sensitive to boson masses $m_\phi\sim\mathcal{O}(0.1~\mathrm{eV})$. For Lorentz-violating interactions, the projected sensitivity exceeds existing constraints across the entire mass range considered and for the best case scenario, the sensitivity for the axion-mediated monopole-dipole coupling at $m_\phi\sim 0.1~\mathrm{eV}$ is about one order of magnitude stronger than the existing individual product bound.

The projected sensitivities to the mixed couplings depend on the choice of several experimental input parameters, including the surface magnetic field of the magnetized sphere, the number of polarized spins in the source slab, the average separation between the two superconducting electrodes, the distance between the source and the superconductor, and the total integration time. Although, the idealized phase shift scales linearly with the observation time, in a realistic Josephson measurement the useful integration time is bounded by the coherence and stability time of the apparatus, beyond which low frequency drifts and other systematics prevent indefinite improvement in sensitivity. For photophilic scalar interactions, the reach is enhanced by larger surface magnetic fields, while for Lorentz-violating scalar and axion-mediated monopole-dipole interactions, it increases with the number of polarized spins in the source.

As a representative benchmark, for the photophilic scalar scenario we find a sensitivity to the mixed coupling $g_{\phi \gamma\gamma}g_{\phi e e}\sim 8\times 10^{-20}~\mathrm{GeV^{-1}}$ at $m_\phi\sim 6\times 10^{-6}~\mathrm{eV}$ for suitable choices of experimental parameters. For Lorentz-violating scalar-mediated interactions, the projected sensitivity reaches $g_{\phi e e}\tilde{J}_e\sim 8\times 10^{-34}$ at $m_\phi\sim 0.1 ~\mathrm{eV}$, exceeding conservative benchmark limits by roughly three orders of magnitude. Similarly, for axion-mediated monopole-dipole interactions, we obtain a sensitivity of $g_{Se}g_{Pe}\sim 10^{-30}$ at $m_\phi\sim 0.1 ~\mathrm{eV}$, representing an improvement of about one order of magnitude over conservative estimates.

Overall, these results demonstrate that JJ interferometry provides a powerful and complementary platform for probing ultralight boson-mediated macroscopic spin-independent and spin-dependent forces, extending laboratory sensitivity into regimes of mass and distance that are inaccessible to conventional mechanical force sensors. We have shown that the Josephson phase can, in principle, respond sensitively to the effective potential experienced by the superconducting condensate, thereby establishing superconducting quantum phase evolution as a precision observable for new physics searches. While the simplified framework considered here is primarily intended to illustrate the underlying mechanism and estimate the attainable sensitivity, a fully realistic experimental realization will require an optimized readout strategy. In this regard, SQUID-based setups provide a natural setting for controlled phase readout and more robust mitigation and instrumental backgrounds. Our analysis therefore serves as a proof-of-principle theoretical foundation and motivates future dedicated SQUID-embedded implementations. 

\section*{Acknowledgments} 
T.K.P would like to thank CERN for their kind hospitality where part of the work has been done. 
D.C. and T.K.P. are supported by the STFC under Grant No.~ST/T001011/1 and ST/X003167/1.
This article is based on the work from COST Actions COSMIC WISPers CA21106 and BridgeQG CA23130, supported by COST (European Cooperation in Science and Technology).

\section*{Data Availability Statement}
This article has no associated data.

\appendix
\section{Response of the superconducting condensate to a scalar-Cooper pair electron interaction}\label{appendix}
The scalar field responsible for the observable Josephson phase shift is assumed to be generated by the external source, namely the magnetized sphere or the polarized slab. In principle, the scalar field can also be sourced by the Cooper pair electron density within the superconducting electrodes through the $g_{\phi e e}$ coupling. However, as we will show in this appendix, the Cooper pair density inside the electrodes is essentially time independent, so the scalar field produced by it generates only a nearly static background potential. In our setup the signal is extracted using lock-in detection at a controlled modulation frequency, implemented either through distance modulation $d(t)$ or magnetic modulation $B_0(t)$. Consequently, any static contribution from the electrode density appears only as a dc offset or slow drift and is rejected by the lock-in readout.

Since, the scalar field couples directly to the Cooper pair electrons of the superconducting condensate, it need not respond as a rigid phase medium. Instead, the scalar-induced perturbation can generate local density fluctuations, which in turn modify the condensate phase dynamics. We describe the superconducting condensate by a macroscopic wavefunction
\begin{equation}
\psi(\mathbf{x},t)=\sqrt{n(\mathbf{x},t)}e^{i\varphi(\mathbf{x},t)},  
\label{app1}
\end{equation}
where $n$ denotes the Cooper pair density and $\varphi$ is the collective condensate phase. In the presence of a scalar-induced potential $V_\phi(\mathbf{x})$, we model the condensate by the effective Schr\"odinger equation  
\begin{equation}
i\partial_t\psi=\Big[-\frac{\nabla^2}{2m_*}+V_\phi(\mathbf{x})+\mu(n)\Big]\psi,
\label{app2}
\end{equation}
where $m_*=2m_e$ is the effective mass of Cooper pair electron and $\mu(n)$ is the chemical potential, which measures how the local energy changes when the local density changes.  

Substituting Eq.~\eqref{app1} into Eq.~\eqref{app2} and separating the real and imaginary parts, we obtain
\begin{equation}
\partial_t n  +\nabla\cdot(n\mathbf{v}_s)=0,~~~ \mathbf{v}_s=\frac{1}{m_*}\nabla\varphi,  
\label{app3}
\end{equation}
which is the continuity equation, where $\mathbf{v}_s$ denotes the superfluid velocity of the Cooper pair condensate and
\begin{equation}
\partial_t\varphi+\frac{(\nabla\varphi)^2}{2m_*}+V_\phi+\mu(n)-\frac{\nabla^2\sqrt{n}}{2m_*\sqrt{n}}=0, 
\label{app4}
\end{equation}
which is the phase equation, respectively. Therefore, the scalar perturbation affects not only the phase, but also the local condensate density. 

Since, the potential is time-independent, we are interested in a static equilibrium configuration inside each isolated superconducting lead. Therefore, 
\begin{equation}
\partial_tn=0, ~~\mathbf{v}_s=0,    
\label{app5}
\end{equation}
which implies $\nabla\varphi=0$. Thus, the phase is spatially uniform within each lead, i.e. $\varphi(\mathbf{x},t)=\varphi_j(t)$. Hence, the phase equation Eq.~\eqref{app4} reduces to
\begin{equation}
\partial_t\varphi_j(t)+V_\phi+\mu(n(\mathbf{x}))-\frac{\nabla^2\sqrt{n}}{2m_*\sqrt{n}}=0, 
\label{app6}
\end{equation}
where the first term depends on time and the remaining terms depend only on space. Therefore, both parts must be equal to a constant $C_j$ within electrode $j$, i.e.,
\begin{equation}
V_\phi+\mu(n(\mathbf{x}))-\frac{\nabla^2\sqrt{n}}{2m_*\sqrt{n}}=C_j,~~~~~~~~~~ \partial_t\varphi_j(t)=-C_j.   
\label{app7}
\end{equation}
This shows that a static, spatially varying scalar potential does not generate a sustained phase gradient within a single electrode; rather, it is screened by a static redistribution of the charge density.

To study the response, we linearize about a homogeneous background,
\begin{equation}
\begin{split}
n(\mathbf{x})\simeq n_0+\delta n(\mathbf{x}), ~~~|\delta n|\ll n_0,\qquad\\
V_\phi(\mathbf{x})=V_{\phi,0}+\delta V_\phi(\mathbf{x}), ~~~|\delta V_\phi|\ll V_{\phi,0},
\end{split}
\end{equation}
where $n_0$ is the background equilibrium density and expand the chemical potential 
\begin{equation}
\mu(n)\simeq \mu(n_0)+\mu^\prime(n_0)\delta n\simeq \mu(n_0)+\frac{1}{\chi}\delta n,    
\end{equation}
at the leading order, where we define
\begin{equation}
\chi\equiv \Big(\frac{\partial \mu}{\partial n}\Big)^{-1}_{n_0},    
\end{equation}
is the static compressibility of the condensate density to the scalar-induced potential. Therefore, in the linearized regime, Eq.~\eqref{app7} reduces to
\begin{equation}
\delta n(\mathbf{x})\simeq -\chi\delta V_\phi(\mathbf{x}),   
\end{equation}
where we remove the spatially constant part and considering $V_\phi$ varies slowly compared to the characteristic length scale so that the quantum pressure can be neglected.

For the two electrodes, we can also write Eq.~\eqref{app7} as
\begin{equation}
\partial_t\varphi_1=-C_1, ~~~~~~~\partial_t\varphi_2=-C_2.    
\end{equation}
Therefore, the Josephson phase difference evolves as
\begin{equation}
\partial_t \Delta \varphi = - (C_1 - C_2).
\end{equation}
To leading order, the constants are given by the spatially averaged scalar potential,
\begin{equation}
C_j \simeq \overline{V_\phi}_j + \text{const},
\qquad
\overline{V_\phi}_j = \frac{1}{V_j}\int_{V_j} V_\phi(\mathbf{x})\, d^3x,
\end{equation}
where $V_j$ is the volume of electrode $j$. Thus,
\begin{equation}
\Delta \varphi(T)
=
-\int_0^T \left(\overline{V_\phi}_1 - \overline{V_\phi}_2\right)\, d\tau.
\end{equation}

This shows that the observable Josephson signal is governed by the difference of the effective scalar potentials averaged over the two electrodes, rather than by the detailed microscopic profile of $V_\phi$ within each lead.

Finally, if the linear size of each electrode is small compared with the scalar field variation length scale, one may approximate
\begin{equation}
\overline{V}_{\phi j}\simeq V_\phi(\mathbf{x}_j),
\label{eq:center_app}
\end{equation}
where $\mathbf{x}_j$ denotes the center of electrode $j$. In this limit, the phase difference reduces to the expressions given in Eq.~\eqref{difft}. Therefore, in the weak coupling regime of interest, the backreaction is subleading and it is sufficient to treat $V_\phi$ as an externally generated perturbation when computing the condensate response.

\bibliographystyle{utphys}
\bibliography{bira}
\end{document}